\documentclass[final,3p,times]{elsarticle}

\usepackage{graphicx,epsfig,amssymb,amsmath,amsthm,amscd,amsfonts,hyperref} 
\usepackage{lineno}
\usepackage{float}
\usepackage{pifont}
\usepackage{stmaryrd}
\usepackage{wasysym}
\usepackage{marvosym}
\usepackage[T1]{fontenc}
\usepackage{yfonts}
\usepackage{textcomp}
\usepackage{lingmacros}
\usepackage[usenames]{color}
\usepackage[mathscr]{eucal}
\usepackage{bm}

\biboptions{sort&compress}

\newcommand{\be}{\begin{equation}}
\newcommand{\en}{\end{equation}}

\makeatletter
\def\ps@pprintTitle{%
  \let\@oddhead\@empty
  \let\@evenhead\@empty
  \def\@oddfoot{\footnotesize{\it To be submitted to: \@journal}\normalsize \hfil\footnotesize{\it\today}}
  \let\@evenfoot\@oddfoot
}
\makeatother

\journal{Mech.\ Res.\ Commun.}

\newcommand{\corrPR}{\textcolor{black}}

\begin{document}
\begin{frontmatter}
\title{Sliding Contact Fraction in Gravity-Driven Dense Cohesionless Granular Flows} 
\author[gpem]{Patrick RICHARD\corref{cor1}}
\cortext[cor1]{Corresponding author.}
\ead{patrick.richard@univ-eiffel.fr}
\author[gpem]{Riccardo ARTONI}
\ead{riccardo.artoni@univ-eiffel.fr}
\address[gpem]{Univ Gustave Eiffel, MAST-GPEM, F-44344 Bouguenais, France}
\author[imft,legi]{Clovis LAMBERT}
\ead{clovis.lambert@univ-grenoble-alpes.fr}
\author[imft]{Rapha\"el MAURIN}
\ead{raphael.maurin@toulouse-inp.fr}
\author[imft]{Pascal FEDE}
\ead{pascal.fede@imft.fr}
\address[imft]{Institut de M\'ecanique des Fluides de Toulouse, Universit\'e de Toulouse, CNRS, Toulouse 31400, France}
\address[legi]{Present address: Univ. Grenoble Alpes, Grenoble-INP, CNRS, LEGI, F-38000 Grenoble, France}

\begin{abstract}
Using large-scale, three-dimensional discrete element simulations, we investigate the kinematic properties of granular chute flows of cohesionless grains, together with the evolution of the coordination number and the fraction of sliding contacts. 
Our results reveal that, under certain conditions, increasing the grain–grain friction coefficient leads to a decrease in the global dissipation of the system. 
This finding highlights that the overall dissipation is not solely governed by the energy lost during sliding contacts, but also by the probability that such contacts occur, a probability that decreases markedly as the friction coefficient increases. 
Together, these effects demonstrate the subtle and nontrivial interplay between grain-scale frictional interactions and macroscopic flow behavior in dense and cohesionless granular materials.
\end{abstract}

\begin{keyword}
granular flows, DEM, sliding contacts, dissipation
\end{keyword}
\end{frontmatter}

\twocolumn
\section{Introduction}\label{sec:Intro}
Among the geometries used to explore the physics of dense and cohesionless granular flows~\cite{Savage1984,GDR_MIDI}, inclined chute configurations are arguably the most extensively studied~\cite{Silbert_PRE_2001,Mitarai_PRL_2005,Bi_2005,Bi_POF_2006,Delannay2007}.  
In such systems, the flow properties are primarily governed by the balance between the gravitational driving force and the frictional resistance exerted by the basal surface (whether flat~\cite{Taberlet2007} or bumpy~\cite{Silbert_PRE_2001}). Owing to this simple but robust force balance, an inclined chute can be seen as a granular rheometer: when a steady state is reached, the gravitational forcing is compensated by basal friction, and the effective friction coefficient of the flow $\mu_{\mathrm{eff}}$ can be determined as $\tan \theta$, where $\theta$ denotes the inclination angle of the chute.

Substantial progress has been achieved in modeling the behavior of dense and cohesionless granular flows~\cite{Berzi_book,Berzi_PRF_2024}, notably through the development of the purely local $\mu(I)$ rheology~\cite{GDR_MIDI}, non-local approaches (\textit{e.g.}~\cite{Bouzid_PRL_2013,Kamrin_PRL_2012,Dsouza_Nott_2020}), and extensions of kinetic theory to dense regimes~\cite{Jenkins2010}.  
Despite these advances, the grain-scale mechanisms governing momentum transfer, dissipation, and the emergence of macroscopic rheology remain under active investigation. In particular, the role of interparticle contacts, especially the transition between sticking and sliding, has yet to be fully elucidated in dense, gravity-driven flows of cohesionless particles.

In this work, we shed light on these questions using discrete element method (DEM) simulations of dense granular flows down an inclined and bumpy plane. Our analysis focuses on microscopic contact dynamics, with a special emphasis on the fraction of sliding contacts and its dependence on flow parameters.  

The paper is organized as follows. Section~\ref{sec:methodo} describes the numerical methodology. Section~\ref{sec:kinematics} reports the evolution of the volume fraction, streamwise velocity, and granular temperature with the investigated parameters and shows that our results are consistent with those reported in the literature. In Sec.~\ref{sec:contacts}, we analyze the interparticle contact network, highlighting the statistics of sliding contacts. Section~\ref{sec:effet_mu} examines how varying the friction coefficient $\mu$ affects both microscopic and macroscopic flow properties. Finally, Sec.~\ref{sec:Conclu} summarizes our main conclusions.

\section{Simulation methodology}\label{sec:methodo}
We employ our own implementation~\cite{Richard_PRL_2008,Richard_PRE_2012} of the classical Discrete Element Method (DEM), which consists in integrating Newton’s equations of motion for a system of $N$ spherical 
 grains that can slightly overlap. This approach requires specifying explicit expressions for the intergranular forces. DEM has been shown to accurately reproduce experimental observations in a wide range of configurations, including gravity-driven flows~\cite{Silbert_PRE_2001,Mitarai_PRL_2005}, sheared granular systems~\cite{Rycroft_PRE_2009}, jammed or near-jammed states~\cite{Majmudar_PRL_2007}, silos~\cite{Yang2024}, and rotating drums~\cite{Richard_SM_2008}. As the method is now well established and extensively documented in the literature, we only detail here the contact-force models used in this study.

For the normal force between two contacting spheres, we use the standard linear spring–dashpot model,
\[
\bm{F_n} = k_n \bm{\delta_n} - \gamma_n \bm{v_n},
\]
where $\bm{\delta_n}$ is the normal overlap vector, $k_n$ the normal spring constant, $\gamma_n$ the normal damping coefficient, and $\bm{v_n}$ the normal component of the relative velocity. The damping term models the dissipation characteristic of granular materials. 

Similarly, the tangential force is modeled as a linear elastic and linear dissipative force in the tangential direction,
\[
\bm{F_t} = -\, k_t \bm{\delta_t} - \gamma_t \bm{v_t},
\]
where $k_t$ is the tangential spring constant, $\bm{\delta_t}$ the tangential displacement, $\gamma_t$ the tangential damping coefficient, and $\bm{v_t}$ the tangential relative velocity at the contact point. The magnitude of the tangential displacement is truncated when necessary to satisfy the Coulomb friction law:
\[
\lvert \bm{F_t} \rvert \leq \mu \lvert \bm{F_n} \rvert ,
\]
where $\mu$ is the grain–grain friction coefficient.

\begin{figure}[htb]
\begin{center}
\includegraphics[width=0.7\columnwidth]{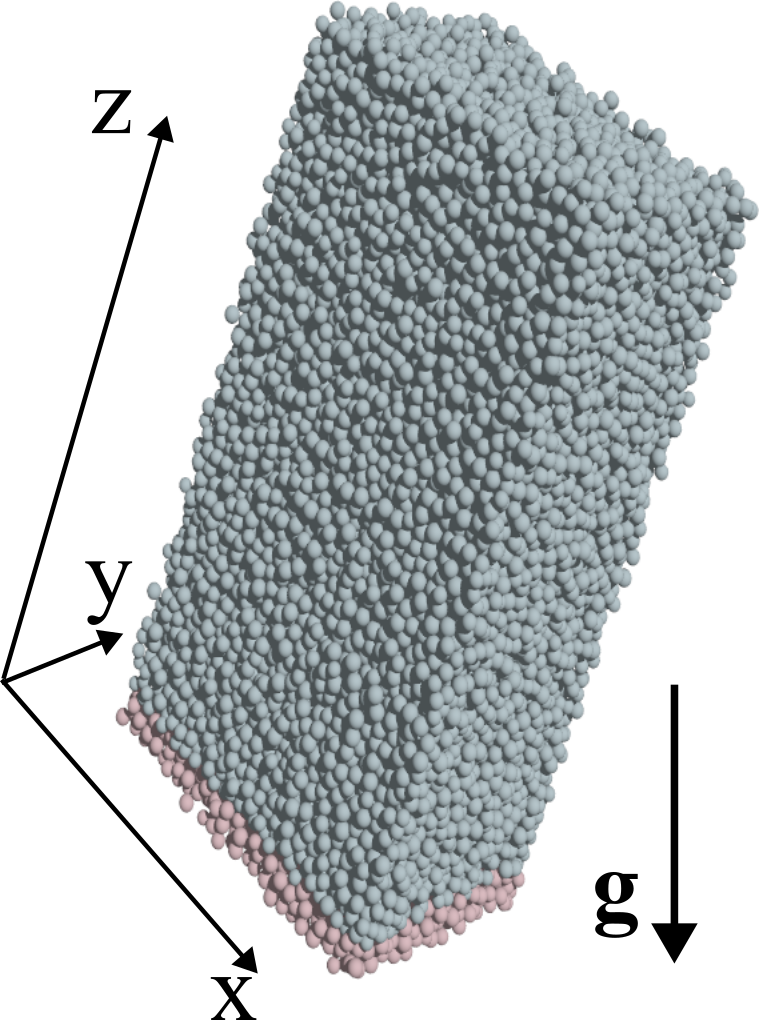}   
\caption{Typical snapshot for chute flow : flow height $H/d=50$ with bumpy bottom surface dimension $25.3d \times 20d$ shown by the pink grains fixed. The inclination angle is $\theta=24^\circ$, the coefficient of restitution is $e_n=0.88$ and friction coefficient $\mu=0.50$. The flow is directed down the incline along the $x$ direction.}\label{fig:sketch}
\end{center}
\end{figure}

We simulate gravity-driven chute flow (Fig.~\ref{fig:sketch}). The grains are slightly polydisperse, with diameters uniformly distributed between $0.9d$ and $1.1d$, and with average mass $m$. The chute is a three-dimensional rectangular box with periodic boundary conditions in the $x$ (flow) and $y$ (vorticity) directions and constrained in the vertical $z$ direction by a basal layer (\textit{i.e.} a fixed bumpy bottom) and a free top surface. The size of the base is $L_X = 25.3d$ in the main flow direction and $L_Y = 20d$ in the vorticity direction. The number of grains $N$ is varied in order to obtain flow heights (defined as $H/d = N d^2/(L_X L_Y)$) between $H/d=5$ and $H/d=50$.

The rectangular box can be inclined by an angle $\theta$ relative to the horizontal (\textit{i.e.}, the angle between the horizontal and the long axis of the base) and is periodic in the $x$ and $y$ directions (see Fig.~\ref{fig:sketch}). The bottom boundary is obtained from the surface of a static granular packing built by setting the inclination to $0^\circ$. More precisely, all grains whose centers lie between the positions of the top layer and those positions minus three mean grain diameters are extracted from the static packing to form the bumpy bottom.

Before any measurements are taken, the inclination is increased to $\theta \approx 40^\circ$, inducing a rapid flow that removes any influence of the initial state. The angle $\theta$ is then reduced to its final value, specified in the captions or in the text below.

We use dimensionless quantities by measuring distances, times, velocities, and elastic constants in units of $d$, $\sqrt{d/g}$, $\sqrt{gd}$ and $mg/d$, respectively. Unless otherwise stated, the parameters used are: $k_n  = k_0  = 5.6\times 10^6 mg/d$, $k_t = {2k_0}/{7}$, $\gamma_t = 0$, and the value of $\gamma_n$ is adjusted to obtain a normal restitution coefficient of $0.88$. Also, unless otherwise specified, the coefficient of friction between grains is $\mu=0.5$.

\begin{figure}[htb]
\begin{center}
\includegraphics[width=0.9\columnwidth]{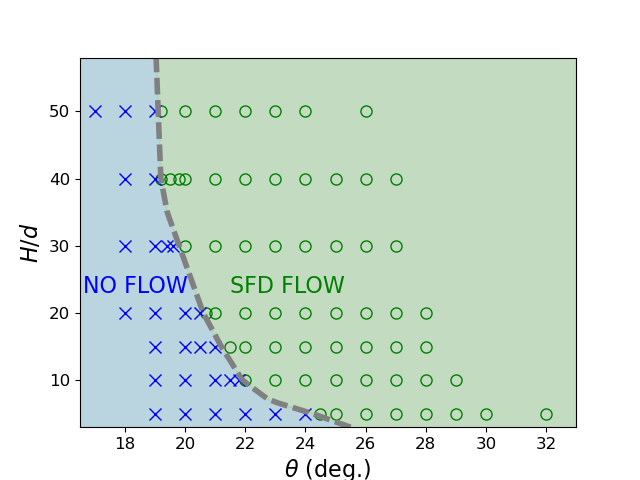}   
\caption{Phase diagram of granular systems in chute flow geometry characterized by the flow height $H/d$ and the inclination angle $\theta$. The bottom is bumpy, the normal coefficient of restitution is $e_n=0.88$ and the friction coefficient is $\mu=0.5$. Empty circles correspond to steady and fully developed flows (SFD flows). Crosses correspond to no sustained flow.}\label{fig:diagphase}
\end{center}
\end{figure}

We draw a phase diagram for the studied flows as a function of the external control parameters: inclination angle $\theta$ and pile height $H/d$ (Fig.~\ref{fig:diagphase}). All simulations performed are represented by markers. For a given flow height $H/d$, below a certain angle (referred to hereafter as $\theta_{\mathrm{stop}}$) no sustained flow is possible. Above this angle, flows can be steady and fully developed (i.e., steady in time and independent of the streamwise direction). The exact location of this boundary depends on model parameters such as the friction coefficient and the restitution coefficient. When the angle is increased further, it has been reported that a shear-thinning layer develops at the bottom of the pile, leading to unstable acceleration of the system~\cite{Silbert_PRE_2001}. We do not focus on this boundary in the present work.

\section{Volume fraction, velocity and granular temperature profiles}\label{sec:kinematics}
In this section, we examine the depthwise profiles of the volume fraction, $\phi$, the streamwise velocity, $V_x$, and the granular temperature, $T$, which are commonly used to characterize the kinematics of granular flows. 
Let us recall that the granular temperature is related to the velocity fluctuations through  
\begin{equation}
T = \frac{1}{3}\left[\langle v_x^2\rangle - \langle v_x\rangle^2 
+ \langle v_y^2\rangle - \langle v_y\rangle^2
+ \langle v_z^2\rangle - \langle v_z\rangle^2\right],
\end{equation}
where $v_i$ denotes the $i-$th component of the velocity and $\langle ... \rangle$ the average operator.
As mentioned above, our analysis is restricted to the steady state. Furthermore, since periodic boundary conditions are applied in the streamwise ($x$) direction, the flow is fully developed, \textit{i.e.}, its properties are independent of the streamwise position, up to statistical fluctuations.

\begin{figure}[htb]
\begin{center}
\includegraphics[width=0.8\columnwidth]{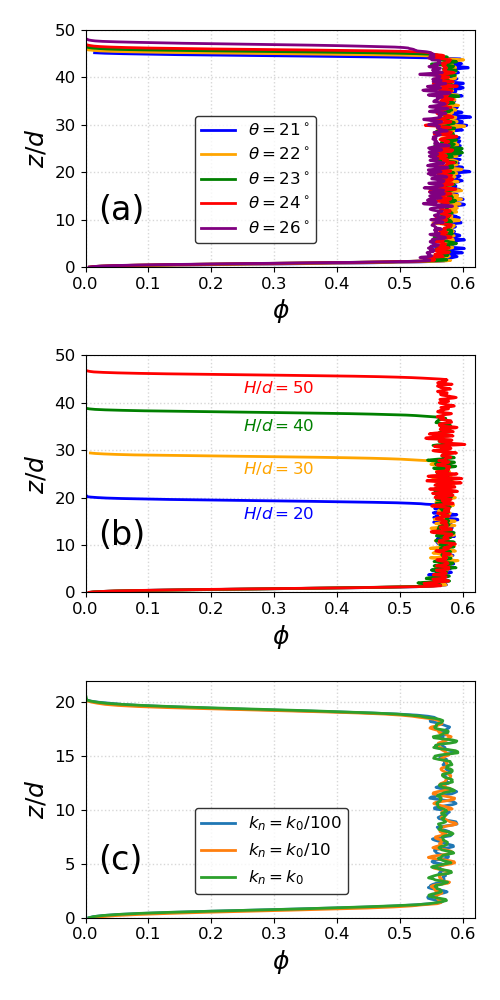}
\caption{Evolution of the depthwise profiles of the volume fraction for (a) several inclination angles ($\theta=21^\circ$, $\theta=22^\circ$, $\theta=23^\circ$, $\theta=24^\circ$, $\theta=26^\circ$) at a fixed height $H/d=50$, (b) $\theta=24^\circ$ and $H/d=20$, $30$, $40$, $50$, and (c) $H/d=20$, $\theta=24^\circ$ and three different values of the grain stiffness, $k_n=k_0/100$, $k_0/10$, and $k_0$ with $k_0 d/mg=5.6\times 10^6$.}\label{fig:compa}
\end{center}
\end{figure}

For the chosen set of parameters, the volume fraction remains nearly constant in the bulk 
over most of the flow depth, with a dilated region appearing only in the vicinity of the free surface (Fig.~\ref{fig:compa}(a)). As the inclination angle $\theta$ increases, the bulk value of the volume fraction decreases slightly but remains in the dense-flow regime, while the thickness of the dilated surface layer increases. These results are in agreement with those reported in~\cite{Silbert_PRE_2001}.

For a given angle, increasing the flow height $H/d$ does not alter the average bulk volume fraction (Fig.~\ref{fig:compa}(b)). The abrupt variations in $\phi$ observed near the bumpy bottom and near the free surface are essentially unchanged when the flow height is varied, indicating that boundary effects are spatially localized within a few grain diameters. Naturally, fluctuations around the mean value appear due to the discrete nature of the grains and the small bin size used when computing the profiles. These fluctuations have a characteristic length scale smaller than one grain diameter and are not associated with any layering phenomenon, which is prevented here by both the bottom bumpiness and the grain-size polydispersity.

We also investigate the influence of the normal and tangential contact stiffnesses, $k_n$ and $k_t$, on the volume fraction profiles (Fig.~\ref{fig:compa}(c)). To this end, we consider three levels of elasticity: $k_n=k_0$, $k_n=k_0/10$, and $k_n=k_0/100$, while keeping the ratio $k_t/k_n = 2/7$ and the coefficient of restitution fixed. The effect of stiffness is found to be weak: the volume fraction is essentially insensitive to the values of $k_n$ explored, within the level of statistical fluctuations. This suggests that, in this regime, the packing density is primarily governed by geometric constraints and shear rate rather than by particle stiffness.

\begin{figure}[htbp]
\begin{center}
\includegraphics[width=0.8\columnwidth]{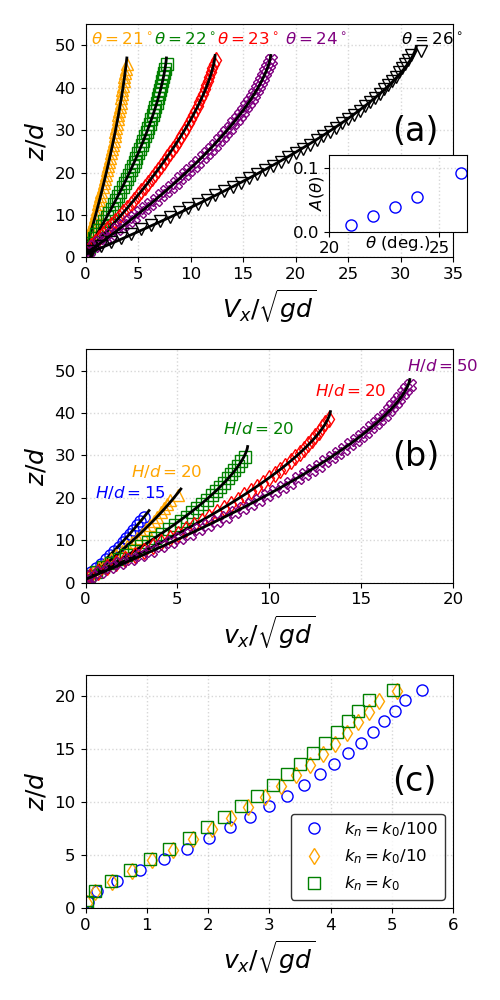}
\caption{Evolution of the depthwise profiles of the streamwise velocity for (a) a height $H/d=50$ and a set of inclination angles ($\theta=21^\circ$, $\theta=22^\circ$, $\theta=23^\circ$, $\theta=24^\circ$, $\theta=26^\circ$), (b) $\theta=24^\circ$ and a set of different flow heights ($H/d=15$, $H/d=20$, $H/d=30$, , $H/d=40$, , $H/d=50$), and (c) $H/d=20$, $\theta=24^\circ$ and three different values of the contact stiffness $k_n=k_0/100$, $k_0/10$ and $k_0$ where $k_0 d / mg =5.6\times 10^6$. 
The numerical profiles are well fitted by the Bagnold velocity profile (see Eq.~\ref{eqn:Bagnold}).
The symbols represent simulations and the solid lines the fit of the numerical data by the Bagnold velocity profile.
The inset of (a) reports the evolution of the fit parameter $A$ of the latter equation versus the inclination angle. 
}\label{fig:vx}
\end{center}
\end{figure}

We report in Fig.~\ref{fig:vx}(a) the depthwise profiles of the streamwise velocity for a set of inclination angles. The profiles are concave, and increasing $\theta$ enhances the average shear rate, resulting in increasingly concave shapes. The evolution of the streamwise velocity profiles with flow height is shown in Fig.~\ref{fig:vx}(b) for $\theta = 24^\circ$.

The vertical velocity profiles are well described by the Bagnold scaling,
\begin{equation}
\frac{V_x}{\sqrt{g\, d}} =  
A(\theta)
\left[
\left(\frac{H}{d}\right)^{3/2} 
- 
\left(\frac{H - z}{d}\right)^{3/2}
\right],\label{eqn:Bagnold}
\end{equation}
where the dependence on~$\theta$ is encapsulated in the coefficient $A(\theta)$, which increases with the inclination angle (see the inset of Fig.~\ref{fig:vx}(a)).
This expression provides an excellent collapse of the data over a wide range of angles of inclination (see Fig.~\ref{fig:vx}(a)) and flow heights (see Fig.~\ref{fig:vx}(b)). It is important to note, however, that the scaling is valid only far from the flow threshold. For $H/d \lesssim 10$, the influence of the bumpy bottom dominates, and the velocity profiles become nearly linear (not shown). This is also true for inclination angles close to $\theta_\mathrm{stop}$, the angle below which, for a given flow height, the flow cannot be sustained. These results are in agreement with the literature~\cite{GDR_MIDI,Silbert_PRE_2001}.

Similarly to the analysis performed for the volume fraction, we have also examined the impact of grain stiffness (Fig.~\ref{fig:vx}(c)). Decreasing the stiffness while keeping the coefficient of restitution constant leads to a systematic increase in the velocity across the depth. The influence of grain stiffness on the velocity profiles is significantly stronger than its effect on the volume fraction, particularly for the least stiff case, which departs markedly from the nominal profile. \corrPR{The results highlight the sensitivity of granular-flow rheology to the micromechanics of contacts, while also revealing the inadequacy of volume fraction as a macroscopic parameter for fully capturing this sensitivity.} \corrPR{Also, in a different geometry, Chialvo et al.~\cite{Chialvo_PRE_2012} have similarly shown that a stiffness-dependence can signal a departure from the inertial regime towards an intermediate flow regime.}

\begin{figure}[htbp]
\begin{center}
\includegraphics[width=0.8\columnwidth]{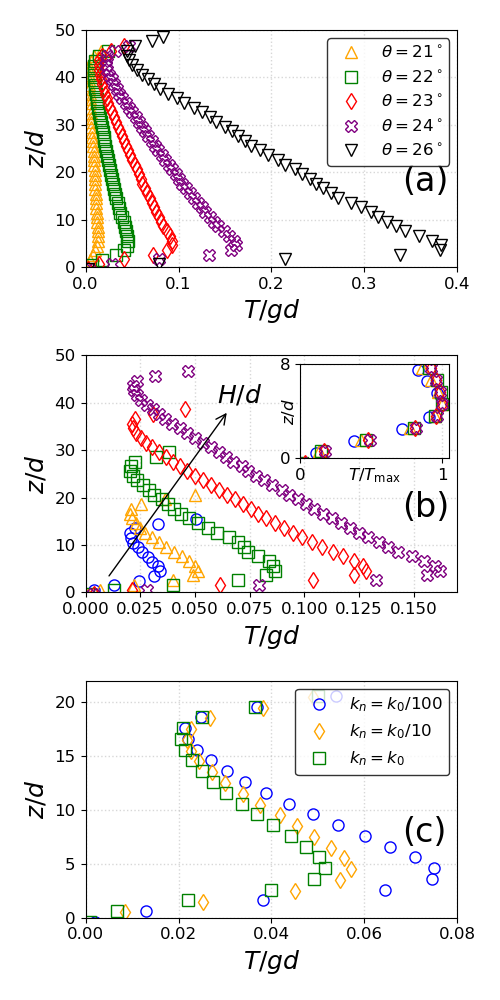}
\caption{Evolution of the depthwise profiles of the granular temperature for (a) several angles of inclination ($\theta=21^\circ$, $\theta=22^\circ$, $\theta=23^\circ$, $\theta=24^\circ$, $\theta=26^\circ$) and a height $H/d=50$, (b) for $\theta=24^\circ$ and a set of flow height ($H/d=15$, $20$, $30$, $40$ and $50$), (c) $H/d=20$, $\theta=24^\circ$ and three different values of the contact stiffness $k_n=k_0/100$, $k_0/10$ and $k_0$ where $k_0 d / mg =5.6\times 10^6$.
The inset of (b) represents the depthwise profiles of the granular temperature normalized by its local  maximal value in the vicinity of the bumpy bottom : $T/T_\mathrm{max}$}\label{fig:T}
\end{center}
\end{figure}

The vertical profiles of the granular temperature for $H/d=50$ and several inclination angles are shown in Fig.~\ref{fig:T}(a). For each inclination, the granular temperature increases from the bumpy bottom, reaching a local maximum a few grain diameters above it. It then decreases as the free surface is approached, before rising again in its immediate vicinity. 
\corrPR{The linear part of the granular temperature profile, and the location of its maximum close to the base, is consistent with the prediction of the purely local $\mu(I)$ rheology. The sharper decay observed within the last few particle diameters above the base suggests an additional, more localized effect of the bumpy bottom on the fluctuating energy budget.}

Increasing the flow height leads to an overall increase in granular temperature (Fig.~\ref{fig:T}(b)). Interestingly, the position of the local maximum near the bumpy bottom is independent of $H/d$ (see the inset of Fig.~\ref{fig:T}(b)), \corrPR{suggesting that this feature is influenced by the bottom bumpiness~\cite{Wang_JFM_2025}.}
In contrast, the height at which the granular temperature starts to rise again near the free surface scales directly with $H/d$, reflecting the geometric constraint associated with the location of the free boundary.

The effect of grain elasticity on the temperature profiles is significant, particularly for low stiffness values. The local maximum close to the bumpy bottom increases markedly as the normal stiffness is decreased (Fig.~\ref{fig:T}(c)). However, one may reasonably expect a saturation at sufficiently large stiffness, where the granular temperature becomes essentially independent of particle elasticity, as the system approaches the rigid-grain limit.

\section{Contact number and fraction of sliding contacts}\label{sec:contacts}
\begin{figure}[htbp]
\begin{center}
\includegraphics[width=0.8\columnwidth]{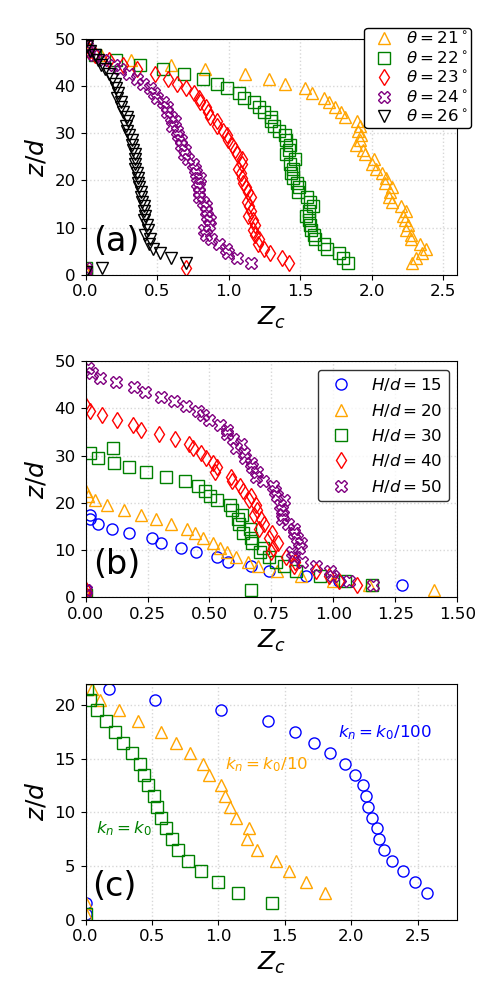}
\caption{The depthwise profiles of the coordination number decrease from the bumpy bottom towards the free surface. 
(a) The profiles increase with the inclination angle (here $H/d = 50$ and $\theta = 21^\circ$, $22^\circ$, $23^\circ$, $24^\circ$, and $26^\circ$). 
For static packings, $Z_c \approx 4.5$ (not shown). 
(b) The profiles also increase with flow height (for a given distance from the bumpy bottom and a fixed inclination $\theta$). 
Here $\theta = 24^\circ$ and $H/d = 15$, $20$, $30$, $40$, and $50$. 
(c) The profiles are strongly influenced by grain stiffness. 
Here $H/d=20$, $\theta = 24^\circ$ and the stiffness values are $k_n = k_0/100$, $k_0/10$, and $k_0$, where $k_0 d / mg = 5.6 \times 10^6$.
}\label{fig:contacts}
\end{center}
\end{figure}

We show in Fig.~\ref{fig:contacts}(a) the depthwise profile of the coordination number $Z_c$, \textit{i.e.}, the average number of contacts per grain, for a set of inclination angles. The flow height is fixed at $H/d = 50$. We observe that the coordination number decreases from the bumpy bottom to the free surface. As expected, higher inclination angles lead to lower coordination numbers. 
The relatively low values of $Z_c$ suggest that, despite the dense nature of the flows, their dynamics are primarily controlled by collisional (inelastic and frictional) interactions, rather than by enduring contacts characteristic of the creeping-flow regime.

Despite the significant flow height, the coordination number does not saturate near the bumpy bottom. However, the influence of the bumpy bottom is clearly visible, especially at large inclination angles. Note that for static packings (not shown), the depthwise coordination profile is essentially constant.

Figure~\ref{fig:contacts}(b) reports the evolution of the depthwise coordination-number profiles for various flow heights. For a given inclination angle and vertical position, increasing the flow height leads to a systematic increase in the coordination number. This trend is observed throughout the packing except in the vicinity of the bumpy bottom, where the influence of flow height is weak or negligible.

We also investigated the effect of grain stiffness on the coordination number. As in the previous analyses, we varied the normal stiffness by considering three stiffnesses: $k_n = k_0$, $k_n = k_0/10$, and $k_n = k_0/100$, while keeping the ratio $k_t/k_n = 2/7$ and the coefficient of restitution fixed. The coordination number is found to depend strongly on grain stiffness: it increases significantly as stiffness decreases. This result is intuitive, since in ``soft-grain'' DEM simulations, grain overlaps become larger as stiffness is reduced, which mechanically increases the coordination number. \corrPR{For low values of $Z_c$ (typically $Z_c < 1$), where contacts can be treated as quasi-instantaneous, independent binary collisions, $Z_c$ scales with $\sqrt{m/k_n}$, consistent with the expected scaling of the collision contact time. For larger $Z_c$, this scaling weakens, reflecting the breakdown of the binary-collision assumption as grains engage in multiple, longer-lasting simultaneous contacts within force chains.}

\begin{figure}[htbp]
\begin{center}
\includegraphics[width=0.8\columnwidth]{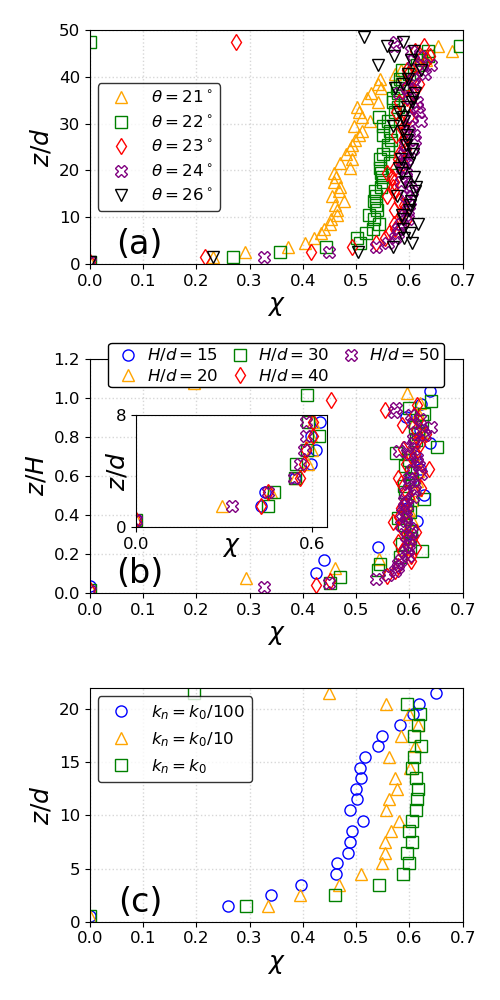}
\caption{(a) The fraction of sliding contacts increases monotonically from the bumpy bottom towards the free surface.  For sufficiently large inclination angles, it reaches an almost constant value (here $H/d = 50$ and $\theta = 21^\circ$, $22^\circ$, $23^\circ$, $24^\circ$, and $26^\circ$). 
(b) The depthwise profiles collapse when $\chi$ is plotted as a function of the rescaled depth $z/H$, except near the bumpy bottom where the variation of this quantity does not depend on $H/d$ (inset of (b)). 
Here $H/d=20$, $\theta = 24^\circ$ and $H/d = 15$, $20$, $30$, $40$, and $50$. 
(c) The fraction of sliding contacts also depends on grain stiffness: reducing the latter leads to a clear decrease in the former. 
Here $\theta = 24^\circ$ and the stiffness values are $k_n = k_0/100$, $k_0/10$, and $k_0$, where $k_0 d / mg = 5.6 \times 10^6$.
}\label{fig:glisse}
\end{center}
\end{figure}

In order to characterize the amount of friction mobilization, we report in Fig.~\ref{fig:glisse}(a) the depthwise profiles of the fraction of sliding contacts, $\chi$, for $H/d = 50$ and several inclination angles $\theta$. At low inclination angles, $\chi$ increases monotonically from the bumpy bottom to the free surface. For sufficiently large angles, however, the profiles become almost flat when approaching the free surface, with a limiting value close to $\chi \simeq 0.6$. This value is consistent with the results reported in~\cite{Silbert_PRL_2002}.

Figure~\ref{fig:glisse}(b) shows the same quantity for a range of flow heights. The flow height does not affect the behavior of $\chi$ near the bumpy bottom (see inset of Fig.~\ref{fig:glisse}(b)). Beyond this region, the depthwise profiles collapse when $\chi$ is plotted as a function of the rescaled depth $z/H$, indicating that the bulk behavior is largely independent of flow height.

As observed previously for the coordination-number profiles, grain stiffness plays a crucial role (Fig.~\ref{fig:glisse}(c)). Decreasing grain stiffness results in a marked reduction in the fraction of sliding contacts. 
A possible explanation may nevertheless be advanced: decreasing grain stiffness enhances interlocking between grains, which can inhibit their sliding.

\section{Effect of the friction coefficient}\label{sec:effet_mu}
In granular materials, energy dissipation arises from several microscopic mechanisms, among which interparticle friction plays a central role. The latter is controlled by the grain–grain friction coefficient $\mu$, which governs both the amplitude of the tangential forces that can be sustained at contacts and the conditions under which contacts transition from sticking to sliding. In this section, we investigate how varying $\mu$ affects both the macroscopic flow properties and the underlying contact dynamics, while keeping all other parameters (grain stiffness and coefficient of restitution) constant.

\subsection{Macroscopic flow properties}

\begin{figure}[htbp]
\begin{center}
\includegraphics[width=0.8\columnwidth]{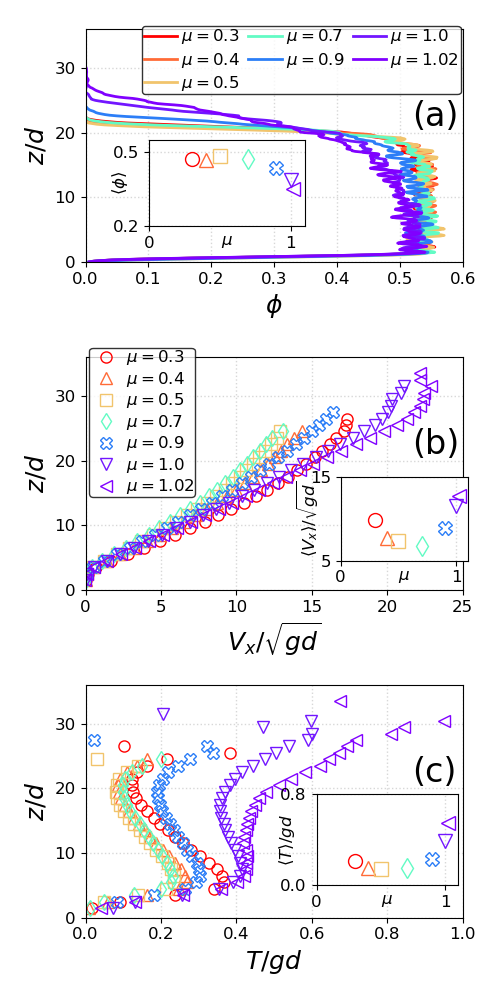}
\caption{(a) The volume fraction depends weakly on the grain–grain friction coefficient up to $\mu=0.5$. Above this value, the volume fraction of the system decreases with increasing friction coefficient. 
(b) The change of behavior is also visible on the velocity profiles which show a deceleration of the system for $\mu\leq 0.5$ then, above this value, an acceleration. (c) Also, the granular temperature decreases with the friction coefficient as long as $\mu\leq 0.5$ then increases significantly. The flow height is $H/d=20$ and the angle of inclination is $\theta=27.5^\circ$.
The insets in (a)--(c) show, respectively, the dependence of the depth-averaged volume fraction $\langle \phi \rangle$, the streamwise velocity $\langle v_x \rangle/\sqrt{gd}$, and the granular temperature $\langle T \rangle/gd$ on the grain--grain friction coefficient $\mu$.
}\label{fig:profil_mu}
\end{center}
\end{figure}

We first examine the influence of $\mu$ on the volume fraction, streamwise velocity, and granular temperature for a representative configuration with flow height $H/d=20$ and $\theta=27.5^\circ$ (Fig.~\ref{fig:profil_mu}).

For friction coefficients in the range $0.3 \leq \mu \leq 0.5$, the depthwise volume fraction profiles remain nearly unchanged (Fig.~\ref{fig:profil_mu}(a)). The flow stays in a dense regime, characterized by a roughly constant bulk volume fraction, with deviations confined to the vicinity of the bumpy bottom and the free surface. In this regime, increasing $\mu$ leads to a progressive reduction of both the streamwise velocity  and the granular temperature (Figs.~\ref{fig:profil_mu}(b) and~\ref{fig:profil_mu}(c) for the profiles and their insets for the average values), consistent with an enhancement of dissipation within the system.
A qualitatively different behavior emerges when the friction coefficient is increased beyond this range. For $\mu \gtrsim 0.6$, the volume fraction decreases throughout the flow depth, indicating a global dilation of the material (see the inset of Fig.~\ref{fig:profil_mu}(a)). Simultaneously, both the streamwise velocity and the granular temperature increase markedly, and even exceed the values observed at lower friction coefficients (see the insets of Figs.~\ref{fig:profil_mu}(b) and~\ref{fig:profil_mu}(c)). This trend is at first sight counterintuitive, as increasing interparticle friction is commonly associated with stronger dissipation and slower flows. These observations suggest that, in this regime, the relationship between microscopic frictional interactions and macroscopic dissipation is non-trivial.



\subsection{Contact network and sliding statistics}

\begin{figure}[htb]
\begin{center}
\includegraphics[width=0.8\columnwidth]{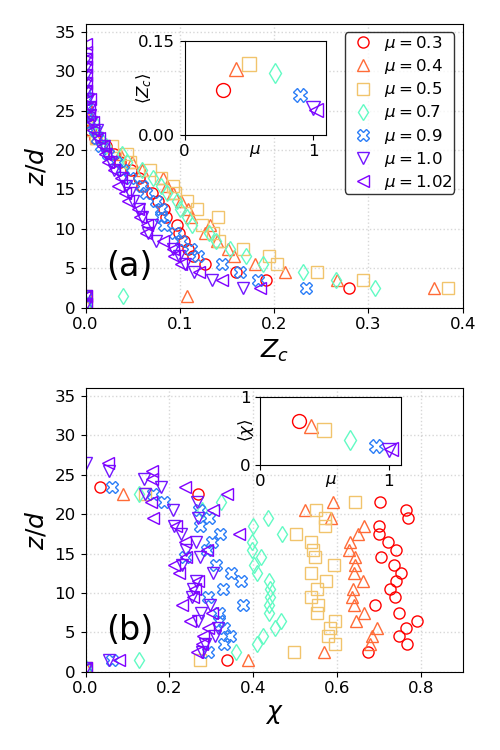}
\caption{(a) For friction coefficients up to $\mu \leq 0.5$, the number of contacts progressively increases with $\mu$. Beyond this threshold, the coordination number decreases. The inset of (a) reports the evolution of the average coordination number versus the grain-grain friction coefficient. (b) The fraction of sliding contacts decreases continuously with $\mu$. The flow height is $H/d=20$ and the angle of inclination is $\theta=27.5^\circ$. The inset of (b) reports the evolution of the average fraction of sliding contacts versus the grain-grain friction coefficient.}\label{fig:Zc_mu}
\end{center}
\end{figure}

To elucidate the origin of this behavior, we analyze the grain-scale contact properties for the same configuration. Figure ~\ref{fig:Zc_mu}(a) shows the depthwise profiles of the coordination number $Z_c$ for different values of $\mu$. For $\mu \leq  0.5$, $Z_c$ increases with $\mu$, indicating a more connected contact network as friction is enhanced. However, beyond this threshold, the coordination number decreases, reflecting a more agitated state with fewer enduring contacts, despite the larger friction coefficient (see inset of Fig.~\ref{fig:Zc_mu}(a)) .

The evolution of the fraction of sliding contacts $\chi$ is shown in Fig.~\ref{fig:Zc_mu}(b) (for the depthwise profiles) and in its inset (for the depth-averaged value). As expected from the Coulomb criterion $\left|F_T\right| \leq \mu \left| F_N \right|$, $\chi$ decreases monotonically with increasing $\mu$. Higher friction coefficients promote sticking contacts and reduce the likelihood of sliding events. This trend holds throughout the flow depth and across the entire range of $\mu$ explored.

\subsection{Implications for frictional dissipation}
The above results indicate that increasing $\mu$ has two competing effects on frictional dissipation. On the one hand, a larger friction coefficient increases the maximum tangential force, and thus the potential energy loss, associated with a single sliding contact. On the other hand, increasing $\mu$ strongly suppresses the occurrence of sliding events (which might be replaced by rolling events~\cite{DeGiuli_PRE_2016}), as quantified by the decrease in $\chi$. \corrPR{Note also that, these observations may indicate that the flows at large friction coefficients progressively depart from the classical inertial regime, although additional investigations would be required to establish the precise nature of this transition.}
\begin{figure}[htb]
\begin{center}
\includegraphics[width=0.8\columnwidth]{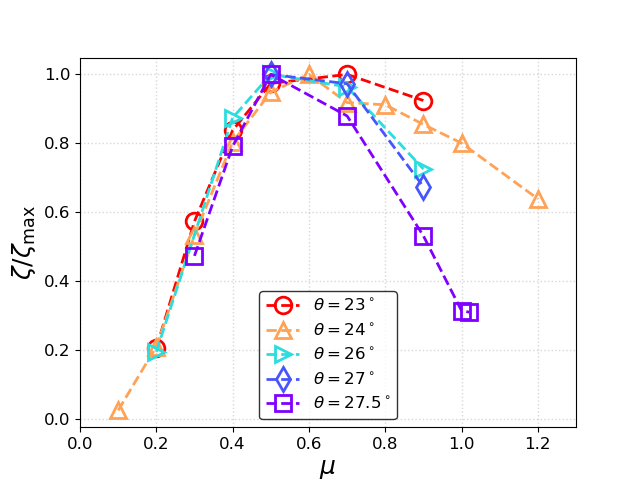}
\caption{With increasing grain--grain friction coefficient $\mu$, the quantity $\zeta=\langle \chi \rangle \langle Z_c \rangle$ (here normalized by its maximal value) exhibits a non-monotonic behavior, reflecting that the decrease in the depth-averaged fraction of sliding contacts dominates over the increase in $\mu$. The quantities $\langle \chi \rangle$ and $\langle Z_c \rangle$ denote the depth-averaged fraction of sliding contacts and coordination number, respectively. The flow height is $H/d = 20$.
}\label{fig:Plateau}
\end{center}
\end{figure}

To capture this competition, we consider the quantity $\zeta =  \langle \chi \rangle \langle Z_c \rangle $, where $\langle \chi \rangle$ and $\langle Z_c \rangle$ denote respectively  the fraction of sliding contacts and the coordination number  averaged over the flow depth. The quantity $\mu\zeta$ \corrPR{provides an indicator of the total mobilized friction}  in the system, and $\zeta$ serves
as a structural indicator of the relative importance of frictional
sliding within the contact network.
Figure~\ref{fig:Plateau} reports the evolution of $\zeta$ normalized by its maximal value as a function of $\mu$ for several inclination angles at fixed flow height $H/d = 20$. 
The dependence of $\zeta$ on $\mu$ is non-monotonic, with an initial increase at low friction followed by a decrease at higher friction suggesting that the reduction in the probability of sliding contacts with increasing $\mu$ might outweighs the increase in dissipation per sliding event. \corrPR{It should be emphasized that the value of $\mu$ at which $\zeta$ begins to decrease likely depends on other parameters, such as the restitution coefficient, which was kept constant throughout this study.}
This decrease in the relative importance of frictional sliding at large $\mu$ is fully consistent with the macroscopic trends reported above and corresponds to the transition  from the \textit{frictional sliding} regime to the \textit{rolling} regime reported in~\cite{DeGiuli_PRE_2016}. In this regime, the flow becomes more agitated, with fewer enduring contacts and enhanced velocity fluctuations, leading to higher velocities and granular temperatures despite the larger friction coefficient. Similar conclusions have been reached in previous studies of frictional granular flows~\cite{Chialvo_PRF_2013,DeGiuli_PRE_2016,Jenkins_2026}, supporting the generality of this mechanism.
Overall, these results highlight that, in dense granular flows, macroscopic dissipation cannot be inferred solely from the value of the grain–grain friction coefficient. Instead, it emerges from a subtle interplay between the strength of frictional interactions and the evolving structure and dynamics of the contact network. \corrPR{Note that increasing the friction coefficient may also favor the transient formation of short-lived force chains that contribute to stress transmission. If such structures are indeed present, they could reduce the contribution of collisional momentum transfer and thereby provide an alternative explanation for the observed decrease in density.}

\section{Conclusion}\label{sec:Conclu}
In this work, we have focused on the steady-state properties of gravity-driven granular flows down a bumpy incline. We performed discrete element method simulations over a wide range of flow heights and inclination angles. Our results reproduce the classical features reported in the literature: a nearly constant volume fraction throughout most of the flow depth, velocity profiles consistent with Bagnold scaling, and a local maximum of granular temperature near the bumpy bottom.

We then examined the contact network in detail and reported the evolution of both the coordination number and the fraction of sliding contacts as functions of the inclination angle, flow height, and grain stiffness. These analyses highlight the sensitivity of the microscopic contact structure to grain-scale mechanical properties and flow configuration.

Finally, we investigated the influence of the grain–grain friction coefficient $\mu$ on the flow rheology. We showed that, although the energy dissipated during a sliding contact increases with $\mu$, the probability that such sliding events occur decreases significantly as $\mu$ grows. As a consequence, in some situations, the effective frictional dissipation associated with sliding contacts decreases with increasing friction coefficient. This counterintuitive behavior underscores the complex interplay between contact dynamics and macroscopic flow properties in dense granular materials.

\section*{Declaration of competing interest}
The authors declare that they have no known competing financial interests or personal relationships that could have appeared to influence the work reported in this paper.

\section*{Acknowledgements}
We thank James T. Jenkins and Dalila Vescovi for fruitful discussions.

\bibliographystyle{unsrt}
\bibliography{biblio.bib}

\begin{thebibliography}{10}

\bibitem{Savage1984}
Stuart~B. Savage.
\newblock The mechanics of rapid granular flows.
\newblock volume~24 of {\em Advances in Applied Mechanics}, pages 289--366.
  Elsevier, 1984.

\bibitem{GDR_MIDI}
GDR MIDI.
\newblock On dense granular flows.
\newblock {\em The European Physical Journal E}, 14:341--365, 2004.

\bibitem{Silbert_PRE_2001}
Leonardo~E. Silbert, Deniz Erta\ifmmode~\mbox{\c{s}}\else \c{s}\fi{}, Gary~S.
  Grest, Thomas~C. Halsey, Dov Levine, and Steven~J. Plimpton.
\newblock Granular flow down an inclined plane: Bagnold scaling and rheology.
\newblock {\em Phys. Rev. E}, 64:051302, Oct 2001.

\bibitem{Mitarai_PRL_2005}
Namiko Mitarai and Hiizu Nakanishi.
\newblock Bagnold scaling, density plateau, and kinetic theory analysis of
  dense granular flow.
\newblock {\em Phys. Rev. Lett.}, 94:128001, Apr 2005.

\bibitem{Bi_2005}
Weitao Bi, Renaud Delannay, Patrick Richard, Nicolas Taberlet, and Alexandre
  Valance.
\newblock Two- and three-dimensional confined granular chute flows:
  experimental and numerical results.
\newblock {\em Journal of Physics: Condensed Matter}, 17(24):S2457, jun 2005.

\bibitem{Bi_POF_2006}
Weitao Bi, Renaud Delannay, Patrick Richard, and Alexandre Valance.
\newblock Experimental study of two-dimensional, monodisperse,
  frictional-collisional granular flows down an inclined chute.
\newblock {\em Physics of Fluids}, 18(12):123302, 12 2006.

\bibitem{Delannay2007}
Renaud Delannay, Michel~Y. Louge, Patrick Richard, Nicolas Taberlet, and
  Alexandre Valance.
\newblock Towards a theoretical picture of dense granular flows down inclines.
\newblock {\em Nature Materials}, 6(2):99--108, Feb 2007.

\bibitem{Taberlet2007}
Nicolas Taberlet, Patrick Richard, James~T. Jenkins, and Renaud Delannay.
\newblock Density inversion in rapid granular flows: the supported regime.
\newblock {\em The European Physical Journal E}, 22(1):17--24, Jan 2007.

\bibitem{Berzi_book}
Diego Berzi.
\newblock {\em With a Grain of Salt}.
\newblock Zenodo, July 2026.

\bibitem{Berzi_PRF_2024}
Diego Berzi.
\newblock On granular flows: From kinetic theory to inertial rheology and
  nonlocal constitutive models.
\newblock {\em Phys. Rev. Fluids}, 9:034304, Mar 2024.

\bibitem{Bouzid_PRL_2013}
Mehdi Bouzid, Martin Trulsson, Philippe Claudin, Eric Cl\'ement, and Bruno
  Andreotti.
\newblock Nonlocal rheology of granular flows across yield conditions.
\newblock {\em Phys. Rev. Lett.}, 111:238301, Dec 2013.

\bibitem{Kamrin_PRL_2012}
Ken Kamrin and Georg Koval.
\newblock Nonlocal constitutive relation for steady granular flow.
\newblock {\em Phys. Rev. Lett.}, 108:178301, Apr 2012.

\bibitem{Dsouza_Nott_2020}
Peter~Varun Dsouza and Prabhu~R. Nott.
\newblock A non-local constitutive model for slow granular flow that
  incorporates dilatancy.
\newblock {\em Journal of Fluid Mechanics}, 888:R3, 2020.

\bibitem{Jenkins2010}
James~T. Jenkins and Diego Berzi.
\newblock Dense inclined flows of inelastic spheres: tests of an extension of
  kinetic theory.
\newblock {\em Granular Matter}, 12(2):151--158, Apr 2010.

\bibitem{Richard_PRL_2008}
P.~Richard, A.~Valance, J.-F. M\'etayer, P.~Sanchez, J.~Crassous, M.~Louge, and
  R.~Delannay.
\newblock Rheology of confined granular flows: Scale invariance, glass
  transition, and friction weakening.
\newblock {\em Phys. Rev. Lett.}, 101:248002, Dec 2008.

\bibitem{Richard_PRE_2012}
Patrick Richard, Sean McNamara, and Merline Tankeo.
\newblock Relevance of numerical simulations to booming sand.
\newblock {\em Phys. Rev. E}, 85, 2012.

\bibitem{Rycroft_PRE_2009}
Chris~H. Rycroft, Ashish~V. Orpe, and Arshad Kudrolli.
\newblock Physical test of a particle simulation model in a sheared granular
  system.
\newblock {\em Phys. Rev. E}, 80:031305, Sep 2009.

\bibitem{Majmudar_PRL_2007}
T.~S. Majmudar, M.~Sperl, S.~Luding, and R.~P. Behringer.
\newblock Jamming transition in granular systems.
\newblock {\em Phys. Rev. Lett.}, 98:058001, Jan 2007.

\bibitem{Yang2024}
Xiaodong Yang, Hui Guo, Lijie Cui, Xiaomin Ding, Kezhen Lv, and Xiaoxing Liu.
\newblock Dem investigation on the dynamic characteristics of wall normal
  stress and particle flow during silo discharge.
\newblock {\em Powder Technology}, 448:120234, 2024.

\bibitem{Richard_SM_2008}
Patrick Richard and Nicolas Taberlet.
\newblock Recent advances in dem simulations of grains in a rotating drum.
\newblock {\em Soft Matter}, 4:1345--1348, 2008.

\bibitem{Chialvo_PRE_2012}
Sebastian Chialvo, Jin Sun, and Sankaran Sundaresan.
\newblock Bridging the rheology of granular flows in three regimes.
\newblock {\em Phys. Rev. E}, 85:021305, Feb 2012.

\bibitem{Wang_JFM_2025}
Teng Wang, Lu~Jing, C.Y. Kwok, Yuri~D. Sobral, Thomas Weinhart, and Anthony~R.
  Thornton.
\newblock Basal layer of granular flow down smooth and rough inclines:
  kinematics, slip laws and rheology.
\newblock {\em Journal of Fluid Mechanics}, 1025:A27, 2025.

\bibitem{Silbert_PRL_2002}
Leonardo~E. Silbert, Deniz Erta\ifmmode~\mbox{\c{s}}\else \c{s}\fi{}, Gary~S.
  Grest, Thomas~C. Halsey, and Dov Levine.
\newblock Analogies between granular jamming and the liquid-glass transition.
\newblock {\em Phys. Rev. E}, 65:051307, May 2002.

\bibitem{DeGiuli_PRE_2016}
E.~DeGiuli, J.~N. McElwaine, and M.~Wyart.
\newblock Phase diagram for inertial granular flows.
\newblock {\em Phys. Rev. E}, 94:012904, Jul 2016.

\bibitem{Chialvo_PRF_2013}
Sebastian Chialvo and Sankaran Sundaresan.
\newblock A modified kinetic theory for frictional granular flows in dense and
  dilute regimes.
\newblock {\em Physics of Fluids}, 25(7):070603, 07 2013.

\bibitem{Jenkins_2026}
James~T. Jenkins and Michele Larcher.
\newblock Unrealistic coefficients of sliding friction in numerical simulations
  of granular materials.
\newblock {\em Granular Matter}, accepted, 2026.

\end{thebibliography}

\end{document}